\documentstyle[epsf]{article}

\newcommand{\be}{\begin{equation}}
\newcommand{\ee}{\end{equation}}

\begin{document}

\begin{flushright}
Liverpool Preprint: LTH 481\\
 \end{flushright}
  
\vspace{5mm}
\begin{center}
{\LARGE \bf The $\eta$ and $\eta'$ mesons in QCD.}\\[10mm] 
{\large\it UKQCD Collaboration}\\[3mm]
 
 {\bf   C. McNeile and  C.~Michael \\ 
Theoretical Physics Division, Dept. of Mathematical Sciences, 
          University of Liverpool, Liverpool L69 3BX, UK }\\[2mm]

\end{center}

\begin{abstract}

 We study the flavour singlet pseudoscalar mesons  from first
principles using lattice QCD. We explore the quark content of the 
$\eta$ and $\eta'$ mesons and  we discuss their decay constants.

\end{abstract}
%


\section{Introduction}

 There is considerable interest in understanding hadronic decays
involving  $\eta$ and $\eta'$ in the final state. The phenomenological
study of hadronic processes involving flavour  singlet pseudoscalar
mesons makes assumptions about their composition. Here we  address the
issue of the nature of the $\eta$ and $\eta'$ from QCD directly,  making
use of lattice techniques. 

 Lattice QCD directly provides a bridge between the underlying quark
description and  the non-perturbative hadrons observed in experiment. 
The amplitudes to create a given meson from the vacuum with a particular
 operator made from quark fields are measurable, an example being  the
determination of $f_{\pi}$. It also allows a quantitative study of the
disconnected quark contributions that arise in the flavour singlet
sector.   The lattice approach  provides other information such as that
obtained by varying the number of quark flavours and their masses. 

In the case of pseudoscalar mesons, the chiral perturbation theory
approach  also provides links between a quark description and the
hadronic states.  For the pion, this has the  well known consequence 
that the  decay constant $f_{\pi}$ describes quantitatively both the
$\mu \nu$  and $\gamma \gamma$ decays. For the flavour singlet
states ($\eta$ and $\eta'$), the situation is more
complicated~\cite{chpt}.  The axial anomaly  now involves a gluonic
component and the definition of decay constants is  not straightforward.
From the chiral perturbation theory description, one expects the mixing
of $\eta$ and $\eta'$ to be most simply  described in a quark model basis.
 In the flavour singlet sector, for pseudoscalar mesons, we then have
contributions  to the mass squared matrix with quark model content $(u
\bar{u} +d \bar{d})/\sqrt{2}$  and $s \bar{s}$ (which we label as $nn$
and $ss$ respectively):

 \be   
   \left( \begin{array}{cc} m_{nn}^2 +2x_{nn} &  \sqrt{2} x_{ns} \\
           \sqrt{2} x_{ns} &  m_{ss}^2+x_{ss}  \end{array} \right)
  \ee

 Here $m$ corresponds to the mass of the flavour non-singlet  eigenstate
and is  the contribution to the mass coming from connected fermion
diagrams while $x$ corresponds to the contribution from disconnected
fermion diagrams. In the limit of no mixing (all $x=0$, the OZI
suppressed case), then we have the quenched QCD  result that the $\eta$
is degenerate with the $\pi$ meson and the $\eta'$ would correspond to
the $s \bar{s}$ pseudoscalar meson. This is not  the case, of course,
and the mixing contributions $x$ are important.

Using as input $m_{nn}$, $m_{ss}$, $m_{\eta}$ and $m_{\eta'}$, the three
mixing  parameters $x$ cannot be fully determined. It is usual to
express the resulting one parameter freedom  in terms of a mixing angle,
here defined by
 \be
 \eta =\eta_{nn} \cos \phi - \eta_{ss} \sin \phi \ \ \ 
 \eta'=\eta_{nn} \sin \phi + \eta_{ss} \cos \phi
 \ee 
 We show the resulting values of the mixing parameters $x$ in the figure
(the  input value for $m_{ss}$ will be discussed later).

 The $\eta$ and  $\eta'$ mesons are often described in an SU(3) 
motivated quark basis, namely
$\eta_8=(u\bar{u}+d\bar{d}-2s\bar{s})/\sqrt{6}$, 
$\eta_1=(u\bar{u}+d\bar{d}+s\bar{s})/\sqrt{3}$.  The mixing angle
$\theta$  in this basis would be given by $\phi-54.7^0$ in a lowest
order chiral perturbation theory. In order to have $f_K \ne f_{\pi}$,
one needs  higher order terms in the chiral perturbation theory
treatment and then the mixing scheme  becomes more
complicated~\cite{chpt} in this basis with more than one angle needed.

 In the SU(3) symmetric limit,  $m_{nn}=m_{ss}=m$ and
$x_{nn}=x_{ns}=x_{ss}=x$, so that  only one mixing parameter is
relevant and the mixing matrix simplifies  considerably to a diagonal
form with elements $m^2$ (octet) and $m^2+3x$ (singlet).  Previous
lattice studies~\cite{kuramashi} have  used degenerate quarks, so have
explored this case and have found that the mixing parameter $x$ is of a
magnitude which  can explain qualitatively the observed splitting
between the $\eta$ and $\eta'$ mesons.

 Here we undertake a  non-perturbative  study in QCD from first
principles which will be able to establish the values of the mixing
parameters $x$, including the pattern of SU(3) breaking.
 This more comprehensive study would take into account the different
masses of the light ($u$ and $d$) quarks and the heavier $s$ quark.
Within the lattice approach, it is not at present feasible to evaluate
using quarks as  light as the nearly massless $u$ and $d$ quarks and
also it is more tractable to use  an even number of degenerate quarks in
the vacuum.  As we shall show, despite these restrictions,  a thorough
study of the mixing between  $\eta$ and $\eta'$ is possible.

 Our lattice study uses  dynamical configurations with $N_f=2$ flavours
of sea quarks of type 1 and we  consider the properties of pseudoscalar
mesons made of either quark 1 or quark 2, where quark 2  corresponds to
a heavier  quark. Thus quark 2 is treated as partially quenched. Here we
have in mind exploring a situation which will  be relevant to treating
strange quark propagation (quark 2) in a vacuum containing only  lighter
quarks (quark 1). 
 We focus here on the results of lattice evaluations, for background to
the methods used see ref.\cite{gusken}.  We address three topics where
lattice input permits us to construct a  firm foundation for the $\eta$,
$\eta'$ mixing:

 \begin{itemize}
 \item From comparing pseudoscalar meson masses with valence quarks of two
different masses (namely meson masses $m_{11}$, $m_{12}$ and $m_{22}$),
we can estimate the  mass $m_{ss}$ of the unmixed $\bar{s} s$ meson,
given the  observed $m_{ns}$ and $m_{nn}$ masses (ie $K$ and $\pi$
respectively).

 \item From measuring the mixing parameters $x_{11}$, $x_{12}$ and
$x_{22}$  between initial and final flavour singlet states  consisting
of either quark 1 or 2 with different masses as above, we can  establish
the pattern of SU(3) breaking in the mixing.

 \item For $N_f=2$ degenerate flavours of quark, we determine  the
pseudoscalar decay constants for the flavour singlet ($P_0$) and
non-singlet ($P_1$) meson. This input allows us to discuss the relation
between  the observed $\gamma \gamma$ decay modes of $\pi^0$, $\eta$ 
and $\eta'$ and the underlying quark content. 

 \end{itemize}

\begin{figure}[h]

\epsfxsize=10cm\epsfbox{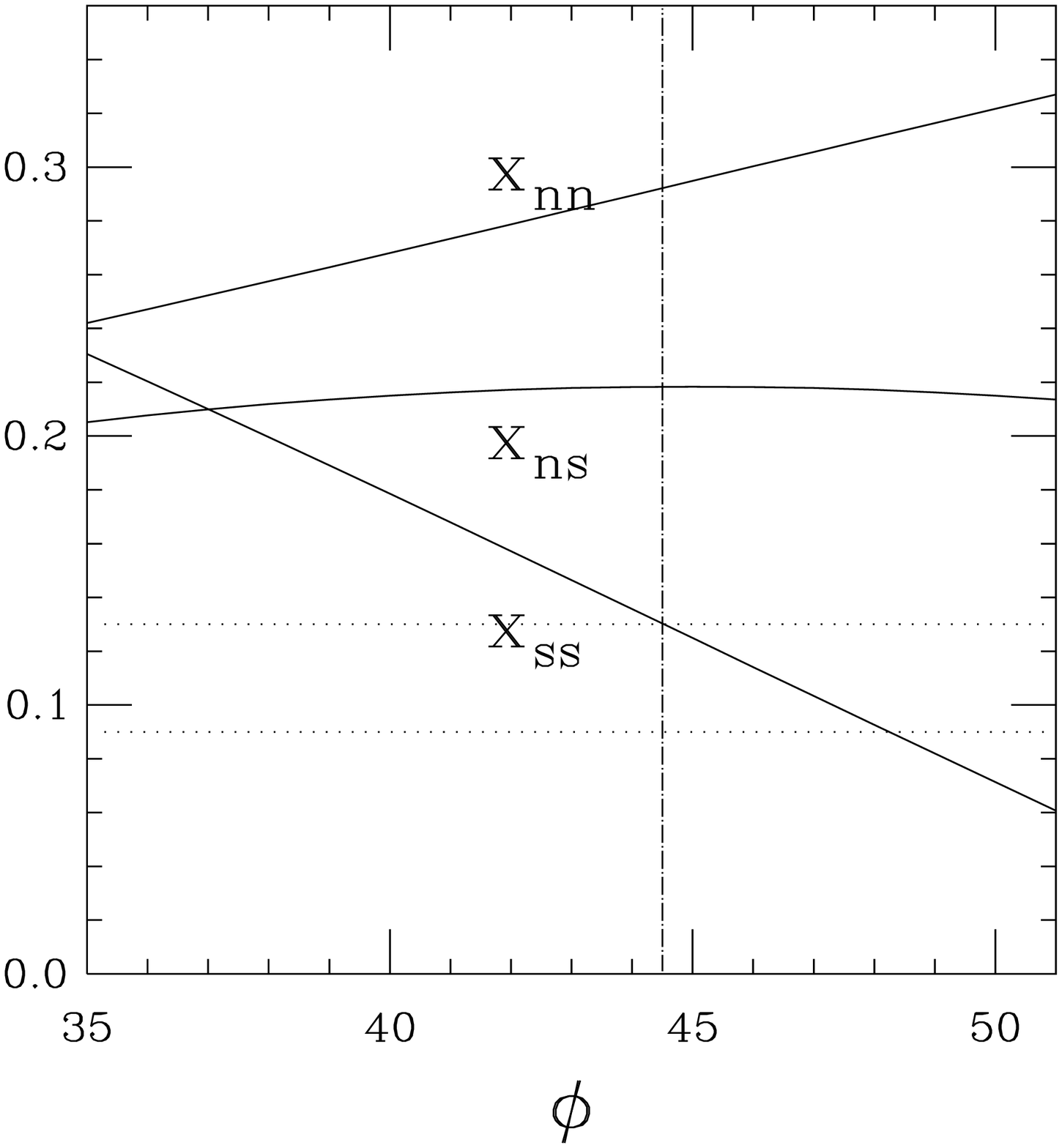}

 \caption{ The mass mixing parameters $x$ in GeV$^2$ versus $\eta$,
$\eta'$ mixing angle $\phi$ in the $\eta_{nn},\ \eta_{ss}$ basis.  The
horizontal dotted lines give the allowed range from the lattice
determination  of $x_{ss}$. The vertical line  illustrates our preferred
solution.
 }
 \label{xxx}

\end{figure}

\section{Lattice results}

 \subsection{ The $s \bar{s}$ pseudoscalar mass}

 Chiral symmetry considerations lead to the expectation that  the
pseudoscalar meson composed of quarks of mass $M_q$ has mass squared
$m^2$  which behaves  linearly with $M_q$ at small quark mass. However,
at large quark mass ($c$ and $b$ quarks for instance), one expects the 
meson mass to vary approximately linearly with the quark mass. Here we
are not concerned with the region of  very small quark mass where chiral
logs are important~\cite{chpt}, so we  summarise this  behaviour by

 \be
   m^2 = b M_q + c M_q^2 + O(M_q^3)
\label{cseq}
 \ee

\noindent  For a pseudoscalar meson  made of two different quarks of
mass  $M_n$ and $M_s$,  we shall assume its  mass only depends on
$(M_n+M_s)/2$ and not on $(M_s-M_n)/2$ as found in lattice
studies~\cite{ukqcd0} and in lowest order chiral perturbation theory. 
 If  eq.~\ref{cseq} were valid with just the linear term  in the quark
mass(ie $c=0$),  then  one directly obtains the required  mass of the
pseudoscalar meson composed of $s$ quarks, $m_{ss}^2=
2m_{sn}^2-m_{nn}^2$, that is $2K^2-\pi^2$, leading to $m_{ss}=0.687 $
GeV.

 This can be explored on a lattice by measuring the pseudoscalar meson
mass  for valence quarks in combinations 11, 22 and 12.  Then, for 
small $c/b$, we have  
 \be
 {c \over 4 b^2} = { {1 \over 2}(m_{11}^2 +m_{22}^2)  - m_{12}^2  \over
       (m_{22}^2 -m_{11}^2 )^2 }
 \ee
 This has been studied in the quenched approximation giving
evidence~\cite{wlat} for a positive coefficient $c$ in eq.~\ref{cseq}.
 In the quenched approximation, however, the chiral behaviour at small
quark mass is anomalous since the theory is not unitary. A better way to
study this issue on the  lattice is then to use dynamical configurations
with sea quarks of type 1 and to  consider the propagation of mesons
made of either quark 1 or quark 2, where quark 2  corresponds to a
heavier  quark. 

 We present results from  UKQCD configurations~\cite{ukqcd} with $N_f=2$
 flavours of sea quark with SW-clover coefficient $C_{SW}=1.76$, lattice
size $12^3 24$, and with sea quarks  having $\kappa=0.1398$,
corresponding to sea quarks of mass around the strange quark mass 
($m_P/m_V=0.67$). Then we  take the heavier valence quark (with
$\kappa=0.1380$) as corresponding to approximately twice the  strange
mass ($m_P/m_V=0.81$). 

The fits with two states to a  $4 \times 4$ matrix of mesonic 
correlators for $t$ range 3 to 10 give results for the spectrum shown in 
Table~\ref{etamass}. 

 Taking account of the correlation among the errors, we obtain the
dimensionless ratio     
 \be
m_{11}^2 c/(4b^2)=0.011(3)
 \ee
  which indicates a statistically significant curvature from the $c$
term. Setting the scale~\cite{ukqcd}  using $a^{-1}=1.47$ GeV then  the
value of $c/b^2$ in physical units can be obtained from $m_{11}=698$
MeV.


Applying  this value of $c$ to the determination of the $m_{ss}$ mass 
from the $\pi$ and K masses,  gives a relative shift upwards due to the
curvature term ($c$) of 1.1(3)\%,  corresponding to a value of
$m_{ss}=0.687+0.008$ GeV. 

Note that this value also helps us to identify the meson mass ratio 
corresponding  to strange quarks, namely
$m_P/m_V=m_{ss}/m_{\phi}=0.682$.

 This lattice study thus answers the question of the  likely deviation
in the pseudoscalar mass formula  from the result given by the lowest
order chiral expression.

\begin{table}[t] 
\begin{tabular}{lllllll} 
\hline 
 $\kappa_s$ & $\kappa_{v1}$ &$\kappa_{v2}$&
  $m_1 a$ & $af_1/Z$ & $m_0 a$&$af_0/Z$ \\
\hline  
0.1398 & 0.1398 & 0.1398& 0.477(5) & 0.171(6) & 0.56(4) &0.196(12)\\
  &&&&&                                   0.56(5) &  0.176(18) \\
0.1398 & 0.1380 & 0.1398& 0.563(5) & 0.182(4)  &  &  \\
0.1398 & 0.1380 & 0.1380& 0.640(4) & 0.190(4) & & \\
\hline

\end{tabular}
 \caption{Pseudoscalar meson masses with $N_f=2$, labelled 1 for
non-singlet (isospin 1)  and 0 for singlet (isospin 0),  and decay
constants for the sea and valence quarks of hopping parameter shown
(here 0.1398 corresponds to strange quarks and 0.138 to  quarks twice as
heavy as strange) from lattice studies with scale $a^{-1}=1.47$ GeV. The
singlet results are shown for  fits with 1 state and $3 \le t \le 7$
(upper) and with 2 states and  $2 \le t \le 7$ (lower).
 \label{etamass}
 }

\end{table}

\subsection{Flavour-singlet mixing}

 The mass splitting between flavour non-singlet and singlet mesons can
be measured using  lattice evaluation of disconnected quark propagators.
This is not an easy task:  the contamination from excited states is
difficult to remove and the  statistical errors turn out to be
relatively large. Initial studies have been in the quenched
approximation~\cite{kuramashi,MQA,gusken}. Here, although there  is no
flavour splitting of the masses,  the mass splitting matrix element $x$ 
can be evaluated.  It is, however, preferable  to be able to study the
mass splitting directly and hence here we  focus on results from full
QCD simulations ~\cite{lat99,sesam}.

The study of the mass spectrum of  flavour singlet ($P_0$) and
non-singlet pseudoscalar meson ($P_1$) using $N_f=2$ flavours of sea
quark 1 leads  to singlet mass  $m_0=(m_{11}^2+2x_{11})^{1/2}$ and
non-singlet mass $m_1=m_{11}$ respectively which allows $x_{11}$ to be
extracted. We shall also be interested in the dependence of $x$ on quark
masses and on non-diagonal mixings. These can be studied  with a little
less rigour as we discuss later.

We use the UKQCD lattices  introduced in the previous section.  The
disconnected  diagrams were evaluated using a variance reduction
method~\cite{lat99} which  uses all the data available with no dilution
from the stochastic method used. We use as many different operators for
the pseudoscalar meson  as possible to  have the largest basis in which
to extract the ground state - local and non-local (fuzzed) in space with
both $\gamma_5$ and $\gamma_5 \gamma_4$ spin structure. Unfortunately, 
even with this  basis of four operators,  we are unable to determine the
singlet mass precisely. For example for both valence and sea quarks
having $\kappa=0.1398$, as shown in Table~\ref{etamass}, we obtain
$am_{0}=0.56(4)$ from  a one state fit to $t=3$ to 7 with a $4 \times 4$
matrix of meson correlators. A two state fit to a wider $t$ range (2-7)
gives a similar mass value.  The corresponding non-singlet pseudoscalar 
mass is also given in Table~\ref{etamass}, so the determination of $x$ 
from $m_1^2+2x_{11}=m_0^2$ has relatively large errors 
($x_{11}=0.10(4)$ GeV$^2$ using $a^{-1}=1.47$ GeV). As an alternative,
we also fit the ratio  of the singlet to non-singlet correlators
directly to a ground state mass difference.  Using the $t$ range 2-7 and
local and fuzzed pseudoscalar operators, we  obtain $am_0-am_1=0.12(7)$.
This method gives a slightly larger mass value (indicating 
$x_{11}=0.13(8)$ GeV$^2$) but  has even larger errors. These results
indicate that  much larger ensembles of gauge configurations will be
needed to make more precise this  approach of  determining $x$ from 
masses.

 If one studies correlations of meson operators made from valence quarks
 of type 2 in a sea of quarks of type 1, one will find the  ground state
 pseudoscalar meson to be that composed of quarks of type 1 (assuming 
type 2 quarks are heavier than type 1). Because of this, we need to 
explore in more detail to study the SU(3) breaking of the mixing
parameters.

 To get a first look at this issue, we consider a quenched lattice and
measure the  ratio of the disconnected to connected diagrams for
pseudoscalar meson propagation. We present  results for $\beta=5.7, \
C_{SW}=1.57,\ 12^3 \times 24$ with 100 configurations with
$\kappa=0.14077$ and 0.13843.  The non-singlet spectrum at these
parameters was studied previously~\cite{shan} giving $m_V/m_P$ values
of  0.65 and  0.78 which  correspond approximately to strange quarks and
quarks twice as heavy as strange. The scale was set as $a^{-1}=1.2$ GeV.
The disconnected  meson correlator was determined using a stochastic
method with variance reduction~\cite{lat99}.

 Assuming  dominance by ground state meson contributions, the ratio
 of disconnected to connected diagrams  at time separation $t$ is 
 \be
{D_{ij} \over C_{ij}}={N_f x_{ij} (t+1) \over  2 (m_{ii} m_{jj})^{1/2} }
 \ee 
 with flavour non-singlet pseudoscalar mass $m_{ii}$ for quarks of type
$i$. The factor of $t+1$ comes from the number of lattice  sites at
which the disconnected diagram can be split. To clarify the pattern of
SU(3) breaking, we also study the  non-diagonal case where we also
measure the disconnected to connected ratio (here $C_{ij}$ is taken as 
$(C_{ii} C_{jj})^{1/2}$). In extracting $x_{12}$ we can make an additional
correction for the contribution from the propagation of mesons with
different masses $m_{11}$ and $m_{22}$  although, in practice,  this
correction is very small.

 \begin{table}[t]
\begin{tabular}{cccc}  
\hline
$t$ &      $x_{11}$ &  $x_{12}$ &  $x_{22}$ \\
\hline
2  &    .089(9) &  .073(6)&   .058(6)\\
3  &   .089(12)  & .072(10) &  .063(10)\\
 \hline \\
\end{tabular}
 \caption{Mass mixing matrix elements $x$ in GeV$^2$ from quenched
lattices for strange valence quarks (1) and quarks of twice the strange
mass (2). }
 \label{xquenched}

 \end{table}

Using local meson operators,  we obtain  for $x$ the values in
Table~\ref{xquenched}. The largest $t$ value has least contributions
from excited state contamination and the consistency of the results
versus $t$ suggests that  such contamination is small.  As was found
previously~\cite{kuramashi}, $x$  increases as the quark mass is
decreased.  Moreover, we can check to see if there is a factorisation 
of $x$ as expected in some chiral perturbation theory
descriptions~\cite{chpt}, namely $x_{12}^2=x_{11} x_{22}$, and we find
that $x_{12}$ lies somewhat below the value given by this assumption.

\begin{table}
\begin{tabular}{cccc}  
\hline
$t$ &      $ x_{11}$ &  $ x_{12}$ &  $ x_{22}$ \\
\hline
2  &    .100(9) &  .072(6)&   .054(5)\\
3  &   .112(12)  & .083(11) &  .063(9)\\
4 &  .106(16)  &   .077(15) &  .059(9) \\
4F &  .093(13)  &   .073(12) &  .052(11) \\
\hline
 \end{tabular}

 \caption{Mass mixing matrix elements $x$ in GeV$^2$ from  lattices with
$N_F=2$ flavours of sea quark (of type 1) with strange valence quarks
(1) and valence quarks of twice the strange mass (2). The  meson
operators used are local except for the case labelled 4F which has
non-local (fuzzed)  creation and annihilation operators  }

 \label{xdf}
 \end{table}

We now revert to discussing the more realistic (partially quenched) 
case: with heavier quarks  of type 2 in a sea of two flavours of quarks of 
type 1. The method described above for the quenched case can be applied
here too. In principle this method is now only valid for small $N_f xt$
for the propagation of quark 1.  From this analysis of the measured 
$D/C$ values, we get the $x$ values shown in Table~\ref{xdf}.
 The values of $ x_{11}$ are similar to those obtained above (with
larger errors) directly  from the rigorous method of using the mass
differences. This suggests that the  strong assumptions made in
determining $x$ directly from $D/C$ are actually  reasonable in
practice. This, and the consistency of values from different $t$ and
different mesonic operators,  gives us confidence to use the $x$ values
from  quarks of type 2 (which are partially quenched anyway) as a guide
to the  quark mass dependence of $x$.   The  $x$ values again show an
increase with decreasing quark mass and also  approximate factorisation.

Setting the quark mass to strange  (since $m_P/m_V=0.682$ in nature for $s$
quarks) in both quenched and  $N_f=2$ evaluations leads to a consistent
lattice estimate of $x_{ss}$ in the range  0.09 to 0.13 GeV$^2$. 
 This value is also consistent with that reported from a study of
$N_f=2$ by the CP-PACS collaboration~\cite{cppacs} with $m_P/m_V=0.69$
and $a^{-1}=1.29$ GeV giving values of   $x_{ss}=0.10$ GeV$^2$ and 0.14
GeV$^2$  (depending on using $t_{\rm min}=2,\ 3$ in fits, respectively).
These lattice values are obtained at quite coarse lattice spacings
 and there may be some additional systematic error  arising from the
extrapolation to the continuum limit. We have, however, chosen to use a 
clover improved fermion action~\cite{ukqcd}  to minimise this
extrapolation error.

 We are unable to determine the mixing strengths $x$ for lighter quarks
than strange.  So we  assume that the value of $x$ continues to
increase as the quark mass is decreased below strange in a similar  way
to the decrease we see from twice strange (type 2) to strange (type 1).

Consider now the consequence of this determination of the mixing. We use
input masses $m_{nn}=0.137$ GeV, $m_{ss}=0.695$ GeV (as discussed above)
and aim to have  $x$ values in line with our results above, namely
$x_{ss} \approx 0.12$ GeV$^2$,   $x_{ns}^2 \approx x_{nn} x_{ss}$ and we
also expect, though with big errors from the extrapolation, 
$x_{nn}/x_{ss} \approx 2$.
 The figure  shows the $x$ values needed to  reproduce the known $\eta$
and $\eta'$ masses for each mixing angle $\phi$.
 The lattice determination of $x_{ss}$ is shown by the dotted horizontal
band.  Keeping close to this band while satisfying the other lattice
constraints  is possible for the mixing illustrated by the vertical
line. This has  $x_{nn}=0.292,\ x_{ns}=0.218, \ x_{ss}=0.13$ GeV$^2$
which gives a description of the  observed $\eta$ and $\eta'$ masses
while being consistent with our  QCD inspired evidence about the mixing
strengths. This assignment  corresponds to a mixing angle $\phi$ in the
$\eta_{nn},\ \eta_{ss}$ basis of 44.5$^0$. Note that this  is almost maximal
which  implies that the quark content (apart from the relative sign)  of
the $\eta$ and $\eta'$ meson is the same. The corresponding mixing angle
in the $\eta_8$, $\eta_1$  basis (modulo comments above) is a value of 
$\theta$ of $-10.2^0$.

\subsection{Flavour-singlet decay constants}

 The decays of $\pi^0,\ \eta$ and $\eta'$ to $ \gamma \gamma$ are
expected to  proceed via the quark triangle diagram. The quark model
gives a decay proportional to $Q_i^2$ for the contribution from a quark of
charge $Q_i$. Thus for the $\pi^0$ meson and the  flavour-singlet $nn$
and $ss$ mesons,  the quark charge contributions to the decay amplitudes
 would be in the ratio  $1:5/3:\sqrt{2}/3$. The experimental~\cite{pdg} 
reduced decay amplitudes for $\pi^0$, $\eta$, and $\eta'$ are in the
ratio $1.0 : 1.00(10): 1.27(7)$.  This information can be used to
analyse the  quark content of the pseudoscalar mesons subject to a
quantitative understanding of the decay mechanisms.

 The conventional approach assumes that the decay constants for the
decays  of the three mesons are the same and then the relative decay
amplitudes give information on the  quark content. This suggests a
mixing angle of $\theta \approx -20^0$ is preferred~\cite{chpt,pdg}.

 We now address the issue of determining these decay constants directly
from QCD using lattice methods.  Our study uses 2 flavours of degenerate
quark  and we define the decay constants by
 \be
 \langle 0 |  A^{\mu} | P_1(q) \rangle = f_1 q^{\mu} \ \ \ \ \ \ 
\langle 0 |  A^{\mu} | P_0(q) \rangle = f_0 q^{\mu}
 \ee

For the isospin 1 state $P_1$ ($\pi$ - like), this is on a firm  footing
because of the anomaly hence $f_1$  will be scale invariant. For the
flavour singlet pseudoscalar meson $P_0$, the decay constant defined as
above will not  be scale invariant because of gluonic contributions to
the anomaly~\cite{chpt}. In this exploratory  study we determine the
decay constants with lattice regularisation and  we shall compare the
singlet and non-singlet values.
 
 These decay constants can be thought of as giving the quark wave
function at the origin of the pseudoscalar meson. Since the mass
splitting  between singlet and non-singlet is not reproduced directly in
quenched QCD, it  is  essential to use lattice studies that  do include
sea  quark effects in this study of decay matrix elements.

 Results were obtained  using fits to full (connected and disconnected)
meson propagation with 4 different types of meson creation and
destruction operator. These are local and fuzzed operators with either
$\gamma_5$ or $\gamma_4 \gamma_5$ couplings, so giving $4 \times 4$
matrix of pseudoscalar  correlators. We used the   $N_F=2$ UKQCD
configurations~\cite{ukqcd} referred to above. For the disconnected
correlators, the variance reduction  technique{~\cite{lat99} is
essential to get a reasonable signal to noise ratio, particularly for
the operators involving the $\gamma_4 \gamma_5$ factor.

The lattice result for $f$ with various valence quark masses with fixed
 sea quark mass (quark 1) as above is shown in  Table~\ref{etamass}. For
the non-singlet results using $a^{-1}=1.47$ GeV and the tadpole-improved
perturbative value of  $Z$  of  0.81 (and of $c_A$ which is
involved in  mixing of the lattice pseudoscalar and axial currents but
has a very small effect in practice) we get  $f_{11}=198(8) $ MeV. Since
this corresponds to strange quarks, it is in reasonable agreement  with
experiment~\cite{pdg} assuming a steady increase from $f_{nn}=131$ MeV
and  $f_{ns}=160$ MeV  to $f_{ss}$. We do see evidence for this increase
in $f$ with quark mass directly on the lattice going from quarks of
type  1 (strange) to type 2 (twice strange) as shown in
Table~\ref{etamass}.

The flavour singlet results are shown in Table~\ref{etamass}. They are
determined by  fits to the appropriate (connected plus disconnected)
meson correlators which  are a  $4 \times 4$ matrix at each $t$ value.
Despite this extensive data set, the determinations of $f$  have
relatively  large statistical errors and the systematic error from
changing the type of fit is also comparable. For our case with $N_f=2$
degenerate quarks, the comparison of the flavour singlet  and
non-singlet shows that the singlet decay constants appear to be somewhat
larger, though the  errors are too big to substantiate this.

 Combining the mass dependence we find in the flavour non-singlet sector
with  the near equality of singlet and non-singlet decay constants, we 
can deduce properties of the physical case with three light quarks. 
Thus,  in terms of the conventional treatment~\cite{pdg}, we would   expect
$f_{\eta}/f_{\pi} > 1$ and $f_{\eta'}/f_{\pi} > 1$. One way to minimise
the effects of mixing is to consider $X =(a_{\eta}^2 +
a_{\eta'}^2)/a_{\pi}^2$ where $a$ refers to the reduced decay amplitude.
Using the conventional formulae for the decay  amplitudes would then
give a value of $X = 3 r^2$ (where $r$ is a suitably weighted
average of  $f_{\eta}/f_{\pi} $ and $f_{\eta'}/f_{\pi}$ which are both
greater than 1). Thus the conventional treatment gives $X >3$ which is 
significantly larger than  the experimental value~\cite{pdg} of
2.64(24).  Thus it appears unlikely that the conventional treatment
(with the decay to $\gamma \gamma$  being given by the analogue of the
formula for pions) is correct for any mixing angle. 
 
 We conclude that there is no support for the conventional assumption
that the singlet decays are given by  a similar expression to the
non-singlet. As has been pointed out by many authors~\cite{chpt}, this 
is  plausible for at least two reasons: (i) the $\eta$ and $\eta'$
mesons are heavier and therefore  less likely to dominate the axial
current or, equivalently, higher order corrections to chiral
perturbation theory will be more important (ii) the flavour-singlet
axial anomaly has a gluonic component which will give additional
contributions to any  hadronic process.

\section{Conclusion}

 From our careful non-perturbative study of mass formulae for flavour
non-singlet pseudoscalar mesons  made of different quarks, we deduce
that the  $ss$ state lies at 695 MeV. We then determine the pattern of
mixing for the  flavour singlet sector, obtaining $x_{ss} \approx 0.12$
GeV$^2$, $x_{nn}/x_{ss} \approx 2$ and  $x_{ns}^2 \approx x_{nn}
x_{ss}$. These conditions are indeed consistent and point to  a  mixing
close to maximal ($\phi=45 \pm 2^0$) in the $nn$, $ss$ basis (this
corresponds to a conventional ($\eta_8$, $\eta_1$) mixing $\theta$  of
$-10 \pm 2^0$). We are able to explore the decay constants  for singlet
pseudoscalar mesons for the first time. Our results show similar  decay
constants for singlet and non-singlet states of the same mass but with
quite large errors. 

 We have not addressed here the issue of the origin of these mixing
parameters $x$.  Lattice studies~\cite{gusken} have the capability to
relate them to topological charge density fluctuations or to other
vacuum properties.

 Our lattice studies have been hampered by two constraints. One is that
the  disconnected quark diagrams needed for a study of singlet mesons
are intrinsically noisy.  Much larger data sets (tens of thousands of
gauge configurations) will be needed  to increase precision. Another
constraint is that we are unable to work with sea quarks  substantially
lighter than strange. We have also not attempted a continuum limit
extrapolation of our lattice results.  Although we are using a lattice
formalism that should improve this extrapolation,  it would be safer to
test it directly. The lattice non-perturbative results do, however, 
show clearly the  structure of the mixing in the singlet pseudoscalar mesons.

\end{document}